\documentclass[aip,apl,reprint]{revtex4-1}
\usepackage{graphicx}
\begin{document}
\title{Spin-orbit field switching of magnetization in ferromagnetic films with perpendicular anisotropy}
\author{D. Wang}
\email{dwwang@nudt.edu.cn}
\affiliation{Department of Physics,
National University of Defense Technology, Changsha 410073, Hunan,
China}
\date{\today}
\begin{abstract}
As an alternative to conventional magnetic field, the effective
spin-orbit field in transition metals, derived from the Rashba
field experienced by itinerant electrons confined in a spatial
inversion asymmetric plane through the \emph{s}-\emph{d} exchange
interaction, is proposed for the manipulation of magnetization.
Magnetization switching in ferromagnetic thin films with
perpendicular magnetocrystalline anisotropy can be achieved by
current induced spin-orbit field, with small in-plane applied
magnetic field. Spin-orbit field induced by current pulses as
short as 10 ps can initiate ultrafast magnetization switching
effectively, with experimentally achievable current densities. The
whole switching process completes in about 100 ps.
\end{abstract}
\maketitle

Ultrafast manipulation of magnetization is currently under intense
investigation, partly driven by the ever increasing demand in
information industry, partly inspired by the intriguing physics
involved. Traditional methods use pulsed magnetic field to realize
ultrafast switching of magnetization, through the spiral motion of
magnetization in a magnetic field, applied in the inverse
direction of the magnetization. However, due to domain wall
instability \cite{kashuba06}, ultrashort field pulses bring about
stochastic behavior, thus imposing limitations on the ultimate
switching speed \cite{tudosa04}. In practice, the limitation on
this switching scheme is related to the difficulty in the
generation of picosecond, strong magnetic field pulses, which
entails the use of relativistic electron bunches nowadays.
Precessional switching scheme, in which the magnetic field is
applied perpendicular to the initial magnetization direction,
circumvents this problem by maximizing the precession torque
experienced by the magnetization \cite{back}. The deficit of
precessional switching is manifested by the needed precise control
of the pulse duration, on the time scale of the magnetization's
precession period. Instead of the conventional magnetic field,
alternative means, such as light \cite{kirilyuk10}, electric field
\cite{ohno00} and electric current \cite{stt}, can be used to
manipulate magnetization. Recently, the effective spin-orbit field
acting on the magnetization attracts much attention because of its
potential applications. This spin-orbit field in transition metals
results from the Rashba field \cite{bychkov84} experienced by
itinerant electrons confined in a spatial inversion asymmetric
potential through the \emph{s}-\emph{d} exchange interaction
\cite{manchon}. Reversible switching of magnetization in
perpendicularly magnetized Co nanodots was already demonstrated
\cite{miron11}, making the speculation of employing the spin-orbit
field to control magnetization in ferromagnetic metals more than
mere imagination, although the underlying mechanism responsible
for the observed switching is still elusive. It is proposed, as
will be shown in the following by macrospin simulation, that the
pure spin-orbit field, in combination with the precessional motion
induced by it, can explain qualitatively the observed experimental
results . In addition, the feasibility of precessional switching
utilizing the spin-orbit field will be addressed as well. It is
found that, due to the large anisotropy and spin-orbit fields,
both derived from the large spin-orbit coupling characteristic of
systems with large perpendicular magnetocrystalline anisotropy
(PMA), the switching time can be as short as 100 ps.

The prototype material system considered here is a trilayer Pt/Co
6 {\AA}/AlO$_x$ nanodot, which is a representative of thin
ferromagnetic metallic nanostructures with PMA. The strong
perpendicular anisotropy results from the 3\emph{d}-5\emph{d}
hybridization at the Pt/Co interface and the 3\emph{d}-2\emph{p}
hybridization at the Co/AlO$_x$ interface \cite{rodmacq09}. The
asymmetry of the top and bottom materials introduces a spin-orbit
field for 3$d$ electrons confined in the thin Co layer. If current
flows along the $x$ direction (c.f. Fig. 2 for the coordinate
system used), and the trilayer structure lies in the $xy$ plane,
then the spin-orbit field is \textbf{B}$_{so}$ = $-\alpha_{so}$
($\hat{\textbf{z}}$ $\times$ \textbf{j}), where $\hat{\textbf{z}}$
is a unit vector along the $z$ axis, \textbf{j} is the current
density, and $\alpha_{so}$ is the spin-orbit field constant, which
is proportional to the spin-orbit coupling in Co. For the
optimized thickness of Co considered here, $\alpha_{so}$ could be
very large, $\alpha_{so}$ = 10$^{-12}$ T m$^2$/A \cite{miron10}.
The induced large spin-orbit field by injecting high density
current into the sample could have profound effects on the
magnetization dynamics.

In the macrospin approximation, the uniform magnetization
$\textbf{M}$ is treated as a macroscopic spin, whose dynamics is
governed by the Landau-Lifshitz-Gilbert (LLG) equation \cite{llg}
\begin{equation}
\label{llg} \frac{d\textbf{m}}{dt} = -\gamma \Big( (\textbf{m}
\times \textbf{B})+ \alpha\, \textbf{m} \times (\textbf{m} \times
\textbf{B})\Big),
\end{equation}
where \textbf{m} = \textbf{M}/$M_s$ is the normalized
magnetization vector ($M_s$ is the magnitude of \textbf{M}),
$\gamma$ = 1.76 $\times$ 10$^{11}$ Hz/T is the free-electron
gyromagnetic ratio, and $\alpha$ is the phenomenological Gilbert
damping constant. The total magnetic field \textbf{B} =
\textbf{B}$_a$ + \textbf{B}$_{app}$ + \textbf{B}$_{so}$ is a sum
of the anisotropy (\textbf{B}$_a$), applied (\textbf{B}$_{app}$)
and spin-orbit (\textbf{B}$_{so}$) fields. In the simulation, the
current flow in Co is along the $x$ axis, while the magnetic field
is applied in the $xz$ plane, 3$^\circ$ tilted away from the $x$
axis. The Gilbert damping is chosen to be $\alpha$ = 0.3
\cite{alpha}. The perpendicular anisotropy field has the form
\textbf{B}$_a$ = $B_Km_z$$\hat{\textbf{z}}$, with $B_K$ = 0.92 T
\cite{miron10}. To stabilize the perpendicular magnetization
configuration, an external field $B_z$ = $\pm$ 5 mT is added to
the total field, depending on the initial magnetization
orientation.

\begin{figure}\centering
\begin{minipage}[c]{\linewidth}
\includegraphics[width=\linewidth]{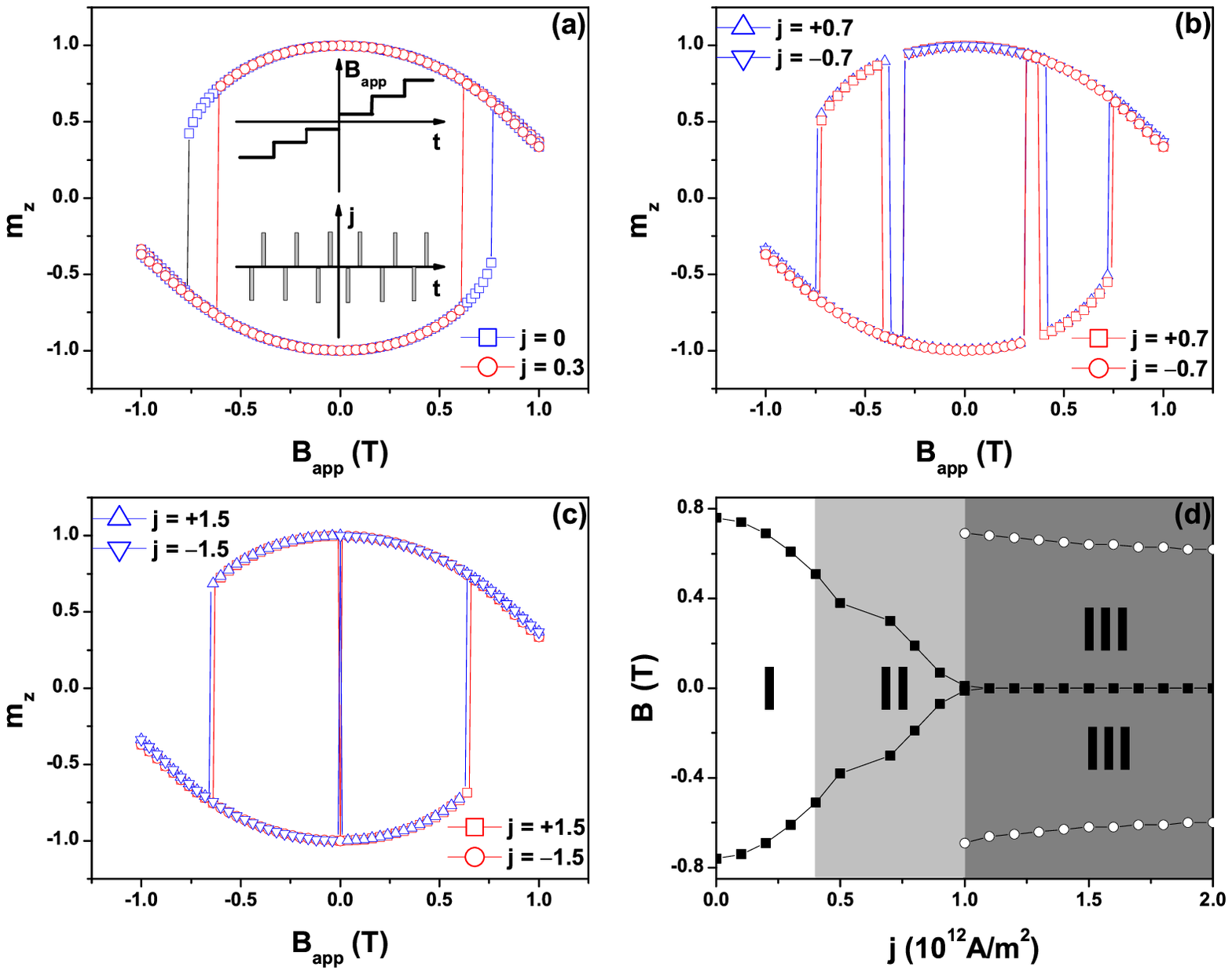}
\end{minipage}
\caption{(a)-(c) Influence of current pulses on magnetization. In
(b) and (c), blue up (down) triangles denote $m_z$ after injection
of a positive (negative) current pulse during the positive to
negative ($+$B $\rightarrow$ $-$B) field sweep, while the
corresponding $m_z$ for the negative to positive ($-$B
$\rightarrow$ $+$B) half is represented by red squares (circles).
The asymmetry between the hysteresis loops for the positive to
negative and the negative to positive field sweeps is caused by
the slightly different paths followed by the magnetization during
the time evolution to equilibrium. Insets in (a) schematically
show the time sequence of the applied magnetic field and current
pulses with both polarities. Current densities are given in units
of 10$^{12}$ A/m$^2$. (d) Dependence of the coercivity (filled
squares) and the maximum field (open circles), against which
current induces magnetization switching, on current density.
Typical hysteretic behavior of regions I, II and III is given in
(a), (b) and (c), respectively.}
\end{figure}

To investigate the effect of the spin-orbit field on the switching
behavior, the time sequence for the applied field and current
pulses, as shown in the insets of Fig. 1(a), is considered.
Essentially, a hysteresis loop is simulated. But at each field
value, current pulses with both polarities, positive and negative,
are applied consecutively. The equilibrium magnetization direction
after each pulse is then recorded. The current pulse is modelled
by a 10 ns square wave with infinitely sharp rising and falling
edges. The spin dynamics under the influence of current is
dictated by the LLG equation, Eq. (\ref{llg}). At the rising edge,
due to the fact that the length of the current pulse is longer
than the characteristic time scale of the magnetization dynamics
triggered by the sudden application of current, the magnetization
stops precessing far before the current pulse is terminated. Once
the falling edge of the current pulse is reached, the
magnetization vector will start precessing again, damping to a
different equilibrium position, depending on the polarity of the
current pulse. The $z$ component of the normalized magnetization,
$m_z$, after positive and negative current pulses, as a function
of the applied field, is shown in Figs. 1(a), 1(b) and 1(c). The
current, or the corresponding spin-orbit field, effect can be
clearly observed: When the current density is lower than 5
$\times$ 10$^{11}$ A/m$^2$ ($B_{so}$ = 0.5 T), only the coercivity
is decreased (Fig. 1(a)). By increasing the current density to
well above 1 $\times$ 10$^{12}$ A/m$^2$ ($B_{so}$ = 1 T),
projection of the magnetization onto the $z$ axis is completely
determined by the polarity of the current (Fig. 1(c)).
Deterministic switching controlled by the polarity of current
occurs. In the intermediate region, current controlled switching
is effective only for a narrow field interval (Fig. 1(b)). Fig.
1(d) gives an overview of the different switching behavior of the
magnetization, for current density ranging from 0 to 2 $\times$
10$^{12}$ A/m$^2$. It can be seen that the coercivity decreases to
zero with increasing current, while the maximum field for current
induced switching remains almost constant, in agreement with
experiment \cite{miron11}.

\begin{figure}\centering
\begin{minipage}[c]{0.8\linewidth}
\includegraphics[width=\linewidth]{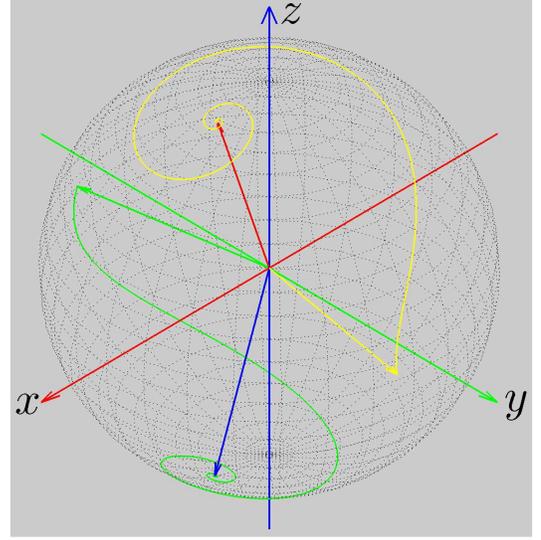}
\end{minipage}
\caption{Three dimensional motion of the normalized magnetization
under the influence of positive and negative current pulses, with
$B_{app}$ = 0.3 T and $j$ = 1.5 $\times$ 10$^{12}$ A/m$^2$. The
red (blue) arrow on the unit sphere represents the equilibrium
orientation of the magnetization after a negative (positive)
pulse, whereas the yellow (green) arrow defines the stable
direction of the magnetization when the current pulse is present.
The yellow (green) curve shows the path of motion for the
magnetization after the negative (positive)  current pulse is
turned off.}
\end{figure}

The physical mechanism responsible for the reversible, current
induced switching can be understood by tracking the magnetization
precession in time. In Fig. 2, two typical precession traces
corresponding to $j$ = $\pm$1.5 $\times$ 10$^{12}$ A/m$^2$ and
$B_{app}$ = 0.3 T are shown. The influence of the polarity of the
current is obvious. It determines whether the magnetization will
spiral upward or downward, initially. If the applied magnetic
field is not too large, which means that the magnetization
orientation pointing up or down is well separated, this initial
discrepancy will lead to the difference in the final equilibrium
position, i.e. whether the magnetization is pointing up or down.
Effectively, the final orientation of the magnetization is defined
by both the spin-orbit field and the applied field, through the
cross product \textbf{B}$_{so}$ $\times$ \textbf{M} \cite{cross},
which is nothing but the initial torque experienced by the
magnetization when the current is turned off. This is consistent
with the symmetry required by the perpendicular switching scheme
\cite{miron11}. In the intermediate region (Fig. 1(b)), the
spin-orbit torque is not large enough to induce switching by
itself, applied field is required to overcome the action of
anisotropy. If the applied field is not large enough, the
equilibrium $m_z$ stays finite even in the presence of the
spin-orbit field, because of the large PMA. When the current is
removed, the magnetization never goes across the $xy$ plane, and
switching could not occur. This explains why the spin-orbit torque
induced switching is effective with large applied field, while
there is no switching if the field is smaller than a critical
value.

When the applied field is rotated away from the $x$ axis by an
angle $\varphi>$  0, the hysteretic behavior of the magnetization
under the influence of current becomes asymmetric, because of the
non-zero $y$ component of the applied magnetic field, $B_{app}
\sin \varphi$. For a positive current pulse to switch the
magnetization, it has to overcome this positive $y$ field, making
the current effect less efficient. The angular dependence of the
maximum field (not shown) against which a positive current pulse
can induce reversible switching supports this intuitive picture.
But, in contrast to the experimental, linear relationship,
theoretically, the dependence is determined by an almost quadratic
relation, $B \propto \cos^2\varphi$. Nevertheless, the overall
decrease in the switching efficiency when the applied field is
rotated away from the current direction is observed unanimously.

\begin{figure}\centering
\begin{minipage}[c]{\linewidth}
\includegraphics[width=\linewidth]{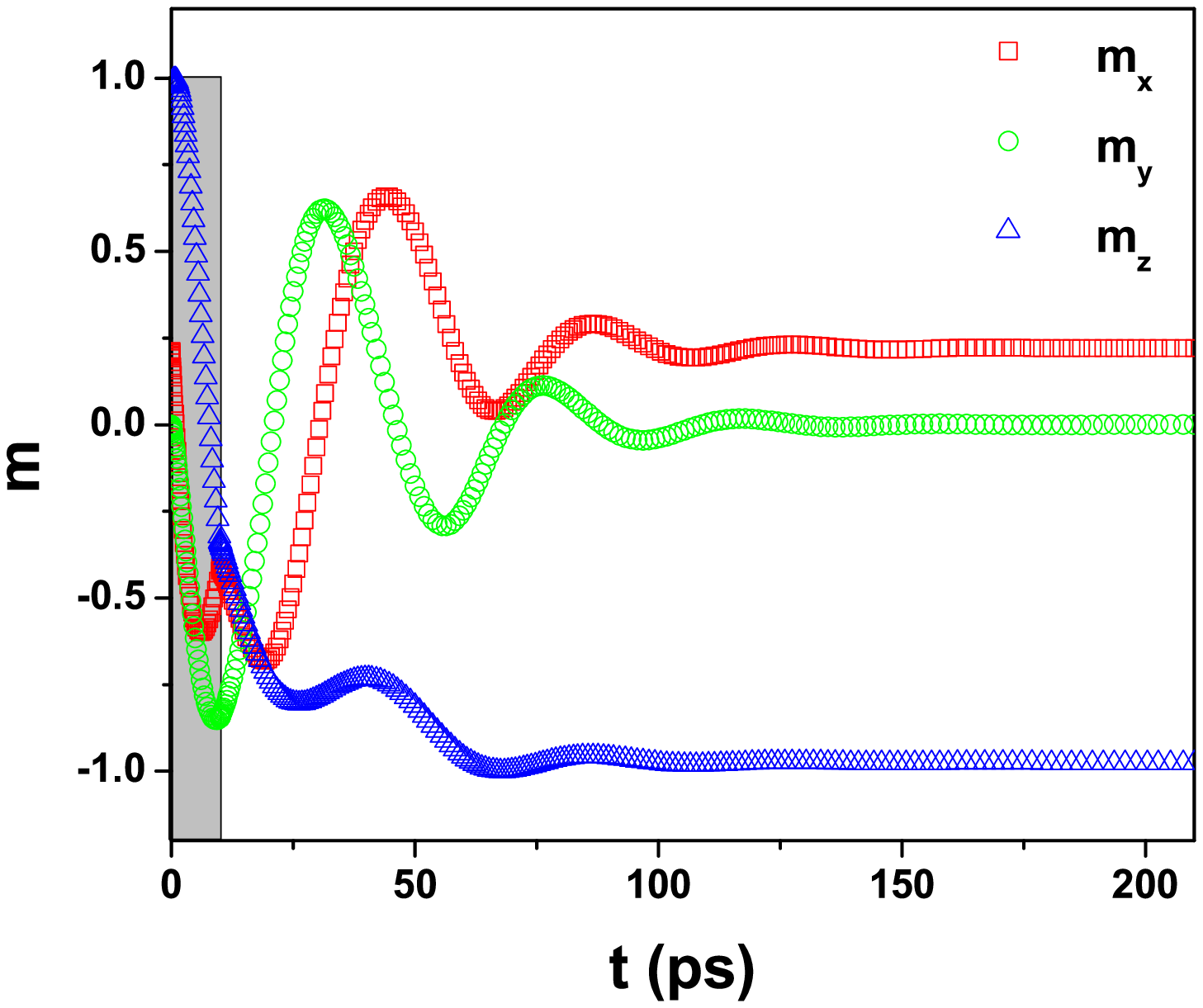}
\end{minipage}
\caption{Time evolution of the normalized magnetization, at
$B_{app}$ = 0.2 T, excited by a 10 ps square-wave current pulse,
whose amplitude is 1.5 $\times$ 10$^{12}$ A/m$^2$. The shaded area
signifies the time interval where the current is present.}
\end{figure}

For the case of current flowing along the $x$ axis, the spin-orbit
field is parallel to the $y$ axis. If the applied field is in the
$xz$ plane, the magnetization is also in the $xz$ plane prior to
the application of current pulses. This perpendicular
configuration between the spin-orbit field and the magnetization
maximizes the precession torque, thus facilitating precessional
switching. Using a square-wave shaped current pulse, complete
switching can be achieved in about 100 ps, as shown in Fig. 3. The
length of the current pulse used in the simulation is 10 ps, which
is about one half of the precession period corresponding to the
spin-orbit field induced by the current pulse, with the current
density $j$ = 1.5 $\times$ 10$^{12}$ A/m$^2$. The applied field is
$B_{app}$ = 0.2 T. Due to the large spin-orbit field, the time
needed to realize precessional switching is solely determined by
the current density, whose direct consequence is the fact that a
very short current pulse can effectively initiate the desired
magnetization switching.

In the macrospin simulation, domain nucleation and the consequent
domain wall motion, which is crucial for the actual determination
of the coercivity, are neglected. Hence the simulated results are
only of qualitative significance. However, as can be seen in Fig.
1, the qualitative agreement between the macrospin simulation and
the experiment \cite{miron11} is satisfactory. Nevertheless, a
detailed micromagnetic study, including finite temperature and
finite size effects, is needed to gain further insight into the
physics involved in the spin-orbit field induced reversible
switching of magnetization in perpendicularly magnetized thin
films. Experimentally, a thorough investigation of the
magnetization dynamics following current excitation in such
systems will prove to be important to clarify the role played by
the spin-orbit field in manipulating the macroscopic state of
magnetization. In Pt/Co/AlO$_x$ or similar systems, this can be
achieved by time resolved magneto optical Kerr effect, which is
already demonstrated to be a powerful technique for the study of
magnetization dynamics in thin metallic magnetic films \cite{van
Kampen02}.

In summary, the spin-orbit field acting on the magnetization,
mediated by the Rashba field experienced by itinerant electrons
confined in a spatial inversion asymmetric plane, through the
\emph{s}-\emph{d} exchange coupling, is proposed for the
manipulation of magnetization. Perpendicular switching of
magnetization in Pt/Co/AlO$_x$ nanodots, with in-plane applied
field, can be realized using only the spin-orbit field, without
the need of any extra fields. This simplifies the explanation for
the experimental observation \cite{miron11}. Ultrafast switching,
on the time scale of 100 ps, is made possible by the large
magnitude of the spin-orbit field in systems with large PMA, such
as Pt/Co/AlO$_x$. For perspectives, the spin-orbit field, properly
tailored, can be used to coherently control spin oscillation and
domain wall motion, in conjunction with the more familiar spin
transfer torques, thus providing more freedom over the control of
magnetization dynamics. The most recent experimental advance on
this respect is the enhancement of domain wall velocity in
perpendicularly magnetized Pt/Co/AlO$_x$ nanowires
\cite{miron11dw}. Stimulated by the impetus from information
technology, more advances are to be expected.
\begin{acknowledgments}
D.W. thanks group Physics of Nanostructures (FNA), Eindhoven
University of Technology for hospitality. Enlightening discussions
with Elena Mure and Sjors Schellekens are acknowledged.
Constructive comments on the manuscript from Zengxiu Zhao and
Jianmin Yuan are gratefully appreciated.
\end{acknowledgments}


\begin{thebibliography}{99}
\bibitem{alpha}
$\alpha$ = 0.3 is close to the experimenally determined Gilbert damping constant
in a similar Pt/Co/AlO$_x$ sample. The detailed determination of
the intrinsic $\alpha$ will be published elsewhere.
\end{thebibliography}
\end{document}